# Utilizing of protein splicing phenomenon for optimization of obtaining and purification of the recombinant human growth hormone

Starokadomskyy* P.L., Okunev O.V., Dubey L.V.

*Institute of Molecular Biology and Genetics, NAS of Ukraine, 150 Zabolotny str., 03143 Kyiv, Ukraine*

**Abstract**

Protein splicing is a post-translational autocatalytic excision of internal protein sequence (intein) with the subsequent ligation of the flanking polypeptides (exteins). The high specificity of excision ensured by intein makes it possible to use a phenomenon of protein splicing for the biotechnology purposes. The aim of this work was optimization of obtaining and purification of the recombinant human growth hormone using the protein splicing. It was experimentally demonstrated that the use of modified intein as auto-removal affine marker makes it possible to perform the rapid and cheap isolation of the recombinant protein Hgh. Furthermore, this approach allows to obtain the human growth hormone with native N-terminus, without formyl-metionine.

Protein splicing is a post-translational autocatalytic excision of internal protein sequence (intein) with the subsequent ligation of the flanking polypeptides (exteins). The high specificity of excision ensured by intein makes it possible to use a phenomenon of protein splicing for the biotechnology purposes (fig.1). The aim of this work was optimization of obtaining and purification of the recombinant human growth hormone using the protein splicing.

The first stage of work was the creation of the chimeric protein, which consists of short N-terminus peptide, intein Mxe GyrA and human growth hormone (Hgh). Theoretical analysis showed that in this chimeric protein intein will ensure the autocatalytic excision of N-terminus peptide and Hgh without their ligation. Actually, we experimentally showed that under certain conditions the chimeric protein effectively splits to form peptides with expected molecular weights. The most effecient process was conducted at room temperature for a period of 4-6 days in the presence of inductor (100 mM β-mercaptoethanol). That allowed to obtain Hgh with native N-terminus (first amino acid Phe, and not formyl-Met).

The next stage of experiment was introduction of affine markers into the body of intein. We tested two types of markers - His-tag (6 histidines) and cellulose-binding domain (CBD). Chimeric protein was purified using affinity chromatography on Ni-sepharose or cellulose, respectively. The subsequent incubation in the presence of β-mercaptoethanol led to scission of the purified chimeric protein. This allows to obtain purified mixture of free intein and Hgh (fig.2). Final purification to separate intein and starting non-scissed chimeric protein was performed by repeated affinity chromatography.

As a result, it was experimentally demonstrated that the use of modified intein as auto-removal affine marker makes it possible to perform the rapid and cheap isolation of the recombinant protein Hgh. Furthermore, this approach allows to obtain the human growth hormone with native N-terminus, without formyl-metionine.

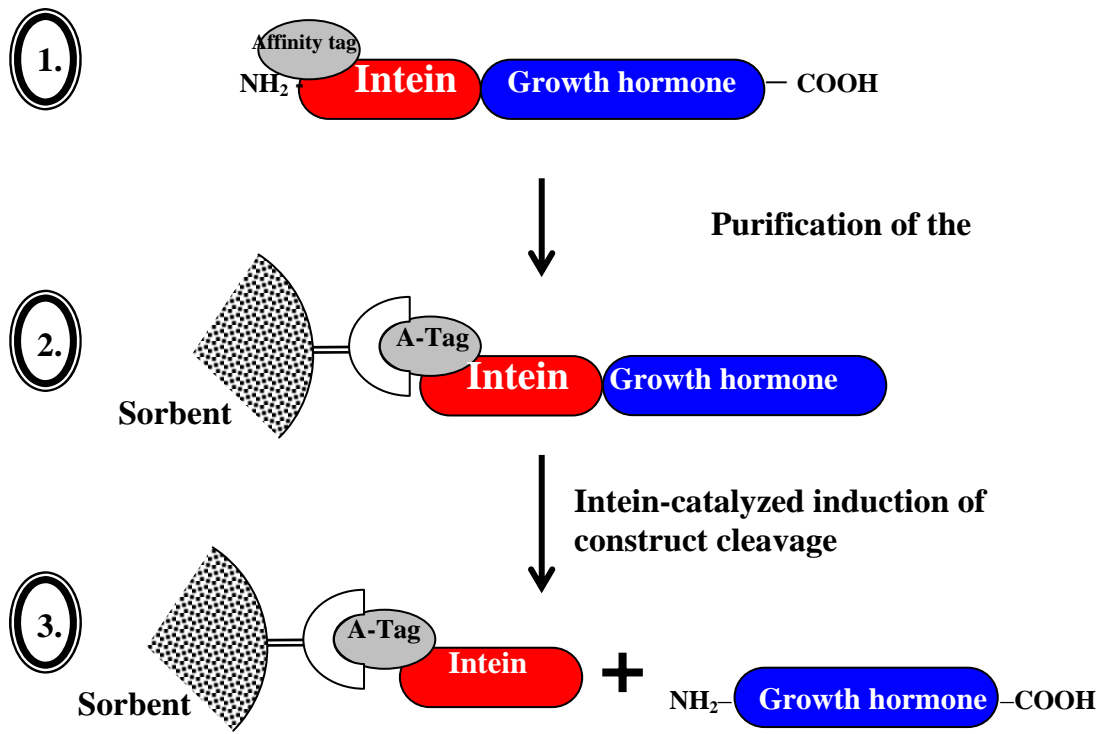

Fig.1. General scheme of protein preparation using protein splicing

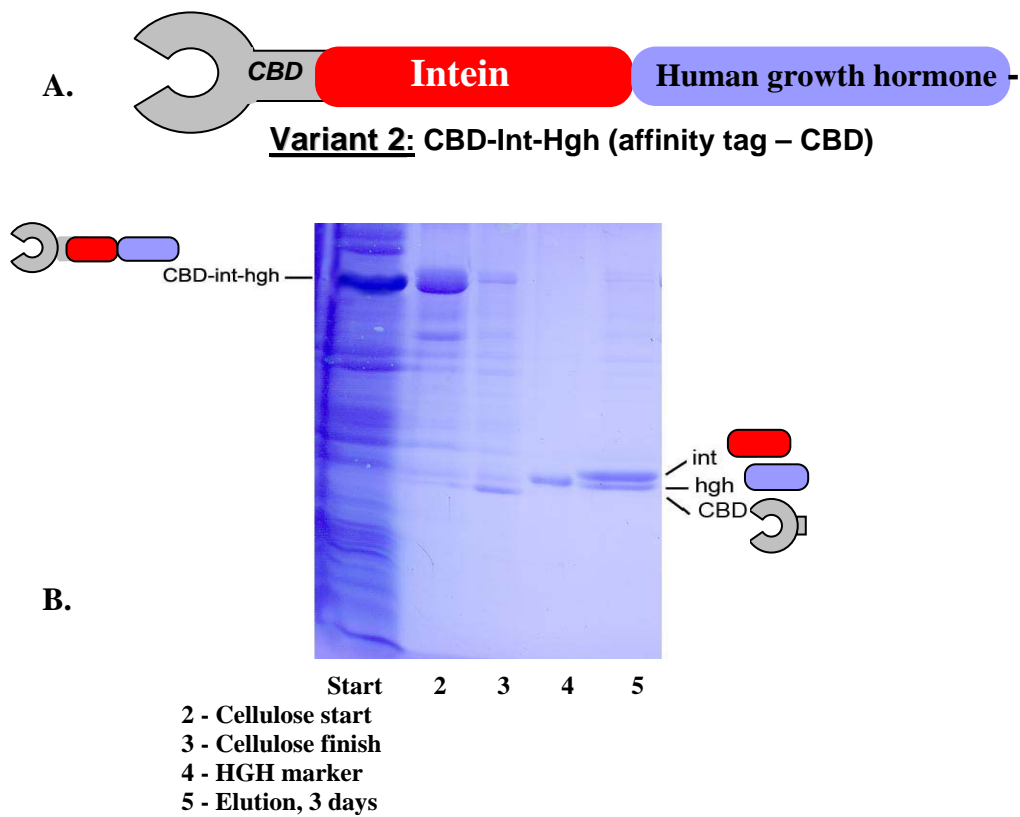

Fig.2. A - Scheme of intein-containing construct; B - PAAG-SDS electrophoresis of the products excision of CBD-Int-Hgh after purification on cellulose